# Exploring Tracer Study Service in Career Center Web Site of Indonesia Higher Education

Renny
Department of Accounting
Gunadarma University
Depok, Indonesia

Reza Chandra
Department of Information Systems
Gunadarma University
Depok, Indonesia

Syamsi Ruhama
Department of Informatics Management
Gunadarma University
Depok, Indonesia

Mochammad Wisuda Sarjono
Department of Information Systems
Gunadarma University
Depok, Indonesia

*Abstract*— Quality competence of worker the present do not meet labor market criteria and the low level of labor productivity, the lack of communication between the labor market with education, changing of socio-economic structure and global political influence labor market, the development of science and technology very rapidly lead to fundamental changes in terms of qualifications, competencies and requirements for entering the workforce. Tracer Study results can be used by universities to determine the success of the educational process that has been done towards their students. Therefore, universities need a technology services to support the optimization of the use of tracer study. One of that is the use of a website to facilitate the conduct tracer study. Most services tracer study provides information to college, like year graduated, got a job waiting period, the first salary to work, first job, the relevance of the curriculum to the work, and compliance with the major areas of work taken in college. Tracer study feature in Career Center Website affect the popularity website especially in traffic and rich file website.

## I. Introduction

The success of the rapid changes in the workplace due to the globalization of the workforce, the revolution in technology and a variety of other disciplines requires anticipation and evaluation of the competencies required by the job.

Evaluation is necessary so that there is no gap between the world of higher education to the world of real work in the community. Some important shifts that occurred include the increase in unemployment of educated, both open and hidden unemployment, as a result of higher education massification, quality competence of workers do not meet labor market criteria and the low level of labor productivity, the lack of communication between the labor market with education, changing of socio-economic structure and global political influence labor market, the development of science and technology very rapidly lead to fundamental changes in terms of qualifications, competencies and requirements for entering the workforce.

How big college graduates are able to act in accordance with the suitability of development education effort to do a search on graduates (Tracer Study). Tracer Study results can be used by universities to determine the success of the educational process that has been done towards their students.

Tracer study is tracking studies trace graduates / alumni conducted between 1-3 years after graduation and study aims to determine the outcome in the form of the transition from the world of higher education to the world of work, output of education is self-assessment against the control and acquisition of competencies, the educational process in the form of an evaluation of the learning process and the contribution of higher education to the acquisition of competencies and educational input in the form of further excavations to information sosiobiografis graduates.

If tracer studies conducted more than 3 years after graduation, tracer studies have several drawbacks, such as the period of retrospection bias due to information that was too far and the information obtained to be less relevant. If done immediately after graduation, the study called exit study, where the study was not able to see the whole process of transition optimally work for a short time after graduating likely unstable work situation or there may even be graduates who have not found a job.

Therefore, we need a technology services to support the optimization of the use of tracer study. One of that is the use of a website to facilitate the conduct tracer study.

Based on this background, the focus of this study is to look at optimizing the utilization of Tracer Study at universities in Indonesia which has a service career center website.

## II. Theoretical Background

Tracer Study is an approach that enables higher education institutions to obtain information about possible deficiencies in the educational process and the learning process and can form the basis for planning activities for the improvement in the future [1].

A tracer study is a graduate or alumni survey that attempts to trace the activities of the graduates or previous students of an educational institution [2].

Tracer study enable the contextualization of graduates of a particular university through a system that is dynamic and reliable in order to determine their life path or movement. It also enables the evaluation of the results of the education and training provided by a particular institution and examines and evaluates the current and future career and employment opportunities / prospects of graduates [3]. Graduates' job titles, years of employment, nature of employment, income levels, and biographical data can be revealed through tracer studies [1].

A tracer study of graduates from the Department of Library and Information Studies at the University of Botswana [4]. The aim of the study was to determine graduates' characteristics, the relevance of their training to their tasks, and their perceptions of the curriculum of the Department of LIS at the University of Botswana. The study revealed that the graduates were employed in traditional library settings. The study also found that their training was relevant to the tasks that they performed, although they advocated the strengthening of the information technology component of the curriculum.

The main objectives of the tracer study were to: investigate the transition process from higher education to: shed light on the course of employment and work over a five year period after graduation; analyse the relationships between higher education and work in a broad perspective which includes the fulfilment of personal goals such as job satisfaction and objective measurement like job position, income, job security and the type of work; find out what factors are important for professional success of graduates taking into account personal factors like gender, work motivation, acquired qualifications during course of study and labour market conditions; evaluate on the basis of the experience and views of graduates, central aspects of the University, including resources, facilities and curriculum and get feedback for their improvement; and identify key aspects of the continuing professional education of graduates, and themes and kinds of courses, including extent, cost, location, reasons for participation, proposals for University courses [5].

In Nigeria a tracer study was done for the Nigerian Teachers' Institute (NTI) which launched its Nigeria Certificate in Education by ODL in 1990 in response to urgent need to train more teachers. The findings of the study were that the performance of ODL graduates was as effective in the classroom as that of their peers who had studied in the traditional way. Their classroom teaching, lesson preparation, motivation of students, record keeping and communication in English was good. The students themselves rated the instructional materials provided quite highly. However the study revealed some dissatisfaction about the use of audio visual material. It was also thought that teachers needed to be better trained in the techniques of ODL . The Institute itself had improved its management and monitoring systems and efforts had been made to address these inadequacies [6].

A higher education institution (HEI) which strives to provide quality education should strive to fully understand the needs of its learners. One of the best ways to do so is through direct feedback from the learners themselves, particularly those who have successfully gone through and completed their study programmes with the institution. Having gone through the system and graduated from it, they are in a very good position to appraise the quality of education that they have received in terms of preparing them to become more holistic individuals in the workplace [7]

## III. Methodology

The sample was 264 universities in Indonesia are included in the ranking of universities based on their activity on the internet, the rank 4ICU and webometrics. Measurements on the first stage is to check whether the college has its own website for the alumni, or herein after called Career Center. Career center is typically an administrative unit of an organization (e.g., school, business, or agency) that employs staff who deliver a variety of career programs and services.

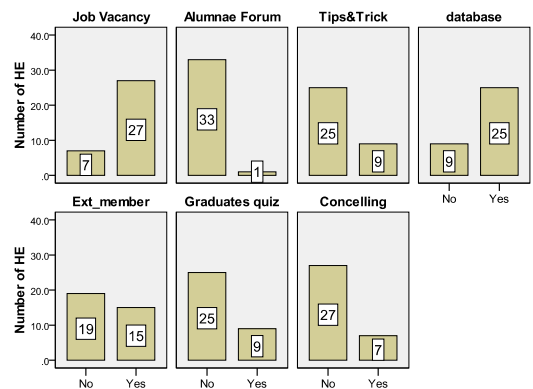

Figure 1. Career Center Website Features

The second stage of 264 samples, which had career center is 34 universities. From 34 universities that have tracer study only 9 universities.The third stage is to examine the features or services that are available in the career website, especially questionnaires alumni.

Observation and measurements conducted research variables in early February 2013. Description of the variables are presented graphically to show utilization the optimization of Tracer Study in universities in Indonesia. With comparing tracer study service in career center website such as curiculum relevance with the implementation in corporate.

## IV. Result and Discussion

### A. Contents

Most services tracer study provides information on the year to go to college, year graduated, got a job waiting period, the first salary to work, first job, the relevance of the curriculum to the work, and compliance with the major areas of work taken in college.

Tracer study feature in Career Center Website affect the popularity website especially in traffic and rich file website.

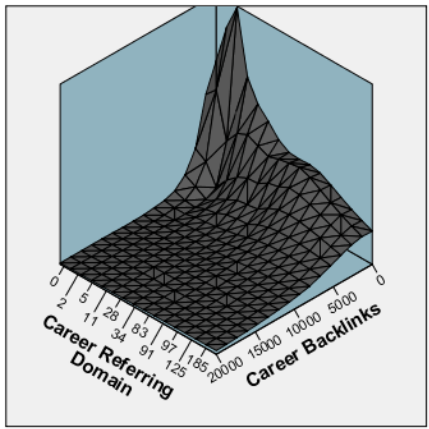

Figure 2. Popularity Career Center Website

Here are 2 examples of web page display face that features a questionnaire tracer or most complete service.

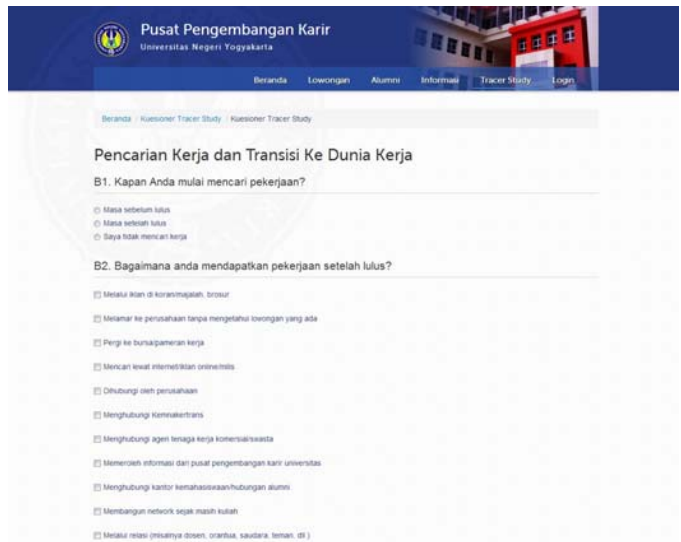

Figure 3. Questionnaires Tracer of Yogyakarta State University (UNY)

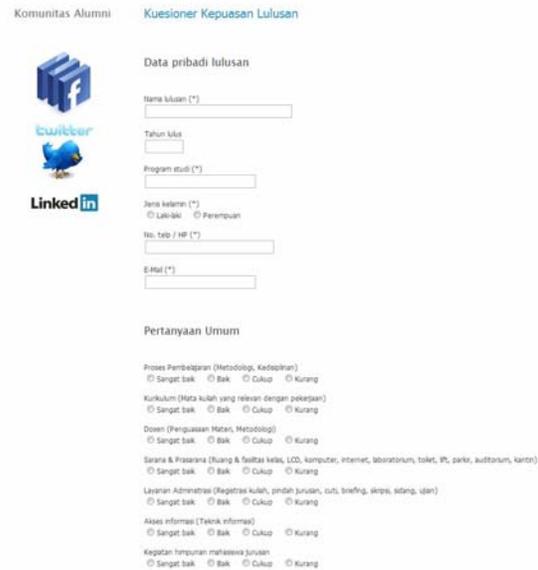

Figure 4. Questionnaires Tracer of Bunda Mulia University (UBM)

Tracer study situation in Indonesia today can be said to still be in the early stages. The analysis conducted by Fikawati and Shafiq [8] indicates that to date information and publications about the tracer study at universities in Indonesia is still very small. The results of the analysis also found that the tracer study in Indonesia vary greatly in terms of clarity of objectives, design and methodology. When compared with the development of tracer study in developed countries, the situation in Indonesia is lagging far enough. In Europe for example, networking tracer study has even produced a large study covering countries in Europe.

## V. Conclusion

Universities in Indonesia have not been optimally utilize the Internet for their graduates. Some information published on the website graduates college, or not on a special site with its own sub domain. Most of the Tracer Study still off line. Variety of features or types of service they provide is still not complete. The most popular type of Tracer Study service is a type of work, the waiting period first job, first salary, entrepreneurship graduates and the relevance of the curriculum.

Questionare that are rarely available is question the number of employees in the workplace, job title, employment status, GPA, how many there are applications for employment sent first.

Tracer Study Service should be focused on the enrichment of question about department needs, such as curriculum, competency, skill (hardskill and softskill).

## *References*

[1] H. Schomburg, Handbook for Graduate Tracer Studies: Centre for Research on Higher Education and Work, Kassel: University of Kassel, 2003.

[2] Millington, "Wikieducator," Open Education Resource Foundation, 1 June 2010. [Online]. Available: http://wikieducator.org/images/e/e1/PID_424.pdf. [Accessed 25 February 2013].

[3] L.O. Aina and K. Moahi, "Tracer study of the Botswana library school graduates," *Education for Information,* vol. 17, no. 3, p. 215, 1999.

[4] L. A. Latif and R. Bahroom, "OUM's Tracer Study: A Testimony to a Quality Open and Distance Education," *ASEAN Journal of Open and Distance Learning,* vol. 2, no. 1, pp. 35-47, 2010.

[5] S. Zembere and N. Chinyama, "Association of African Universities (AAU)," 1996. [Online]. Available: http://www.aau.org/studyprogram/notpub/ZEMBERE.pdf. [Accessed 25 February 2013].

[6] N. Boaduo, J. Mensah and S. Babitseng, "Tracer study as a paradigm for the enhancement of quality course programme development in higher education institutions in South Africa," University of North-West, Potchefstroom, 2009.

[7] U. Abdurrahman, "Quality Assurance Procedures in Teacher Education: The case of the National Teachers," in *Towards a Culture of Quality*, Vancouver, Commonwealth of Learning, 2006, pp. 73-84.

[8] S. Fikawati and A. Syafiq, "Tracer Study Report," CDC-UI, Depok, 2008.